\documentclass[11pt,a4paper]{article}

\usepackage{amsmath}
\usepackage{graphicx}
\usepackage{amsfonts}
\usepackage{amssymb}

\begin{document}
\hfill\hbox{SISSA 72/2006/EP}

\bigskip

\begin{center}
{\Large \textbf{$\boldsymbol{N}\mathbf{{=}1}$ SUSY Conformal Block Recursive
Relations}}
\end{center}

\begin{center}
\vskip 1.0cm {\large
V.~A.~Belavin${}^{a,b}$}\\

\vspace{.4cm}
{\it
$^a$ Institute for Theoretical and Experimental Physics (ITEP)\\
B.~Cheremushkinskaya 25, 117259 Moscow, Russia

\vspace{0.3cm}
$^b$ International School for Advanced Studies (SISSA) \\
Via Beirut 2-4, 34014 Trieste, Italy \\
INFN sezione di Trieste \\
e-mail: belavinv@sissa.it
}
\end{center}

\begin{abstract}
We present explicit recursive relations for the four-point superconformal
block functions that are essentially particular contributions of the given
conformal class to the four--point correlation function. The approach is
based on the analytic properties of the superconformal blocks as functions
of the conformal dimensions and the central charge of the superconformal
algebra. The results are compared with the explicit analytic expressions
obtained for special parameter values corresponding to the truncated operator
product expansion. These recursive relations are an efficient tool for
numerically studying the four--point correlation function in Super Conformal
Field Theory in the framework of the bootstrap approach, similar to that in
the case of the purely conformal symmetry.
\end{abstract}

\section{Introduction}

Recent progress in Liouville field theory~\cite{ZZ1,DO,T1} and
two-dimensional quantum Liouville gravity~\cite{HEM,3tMLG,4tMLG2} implies new
applications in string theory. A very important step in this direction is the
supersymmetric extension of the methods used in the bosonic case. This is
especially interesting because there is not yet a supersymmetric
generalization of the matrix model technique; the super-Liouville approach
introduced by Polyakov~\cite{Polyakov} therefore remains the only promising
approach. One of the open problems here is to construct the complete set of
explicit correlation functions in the minimal supergravity (minimal
superstring theory). As in the bosonic conformal field theory, the main
method here is based on solving the conformal bootstrap equations. The
conformal block functions~\cite{BPZ} play an important role in this program.
Unfortunately, closed analytic expressions for these functions could be found
only for some special values of the conformal dimensions of the fields under
consideration. We here present recursive relations for the four-point
conformal block functions in the Neveu--Schwarz sector of the $N{=}1$ SUSY
conformal field theory analogous to those found in~\cite{AlZ1} in the bosonic
case. Successive iterations of these relations converge for sufficiently
small $x$ and allow efficiently calculating series expansions for the
correlation functions.

The paper is arranged as follows. In Sec.~2, we very briefly recall some
necessary facts about the $N{=}1$ SUSY conformal field theory
(see~\cite{Pogosian,Alvarez} for more details). In Sec.~3, we introduce the
superconformal block functions and describe their analytic properties. In
Sec.~4, we describe the singularities of the superconformal blocks as
functions of the field dimensions. In Sec.~5, we discuss the asymptotic behavior of the conformal blocks as
functions of the central charge of the superconformal algebra $c$ and their
singularities in $c$. We then introduce the recursive relations for the
superconformal block functions. In Sec.~6, we discuss a special degenerate
case resulting in the ordinary differential equations for the superconformal
block functions under consideration and verify the recursive relations for
the corresponding parameter choices. We present our conclusions in Sec.~7.

\section{$\boldsymbol{N}\mathbf{{=}1}$ superconformal field theory}

The symmetry in the superconformal field theory is generated by the
holomorphic and antiholomorphic components of the supercurrent $S$ and the
stress tensor $T$. In terms of the Laurent components of $S$ and $T$, the
algebra takes the conventional form of the Neveu--Schwarz--Ramond (NSR)
algebra
\begin{equation}\label{The algebra}
\begin{aligned}
\left[L_n;L_m\right]&=(n-m)L_{n+m}+\frac{c}{8}(n^3-n)\delta_{n;-m},
\\
\{G_r;G_s\}&=2L_{r+s}+\frac{1}{2}c\left(r^2-\frac{1}{4}\right)\delta_{r;-s},
\\
[L_n;G_r]&=\left(\frac{1}{2}n-r\right)G_{n+r},
\end{aligned}
\end{equation}
where
\begin{equation*}
\begin{alignedat}{2}
&r,s\:\in\mathbb{Z}+\frac{1}{2}&\quad&\text{for NS sector},\\
&r,s\:\in\mathbb{Z}&\quad&\text{for R sector}.
\end{alignedat}
\end{equation*}
In a Liouville-like manner, we write the central charge
\begin{equation}
c=1+2Q^2,\quad\text{where }Q=b+\frac{1}{b}.
\end{equation}
Local fields form the highest-weight representations of the NSR algebra. Each
representation $[\Phi_{\Delta}]$ consists of a primary field with the
conformal dimension $\Delta$ and all its superconformal descendants. The
primary NS superfields\footnote{For simplicity, we consider only the
holomorphic part.} are
\begin{equation}\label{superfield}
\mathbf{\Phi}_{\Delta}(z)=\Phi_{\Delta}(x)+\theta\Psi_{\Delta}(x)\quad
\text{with }\Psi_{\Delta}(z)=G_{-\frac{1}{2}}\Phi_{\Delta}(z),
\end{equation}
where $x$ and $\theta$ are the holomorphic coordinates of the
(2$+$2)-dimensional superspace ($\theta$ is the anticommuting, ``odd"
coordinate). We also introduce the convenient parameterization
\begin{equation}
\Delta(\lambda)=\frac{Q^2}{8}-\frac{\lambda^2}{2}.
\end{equation}
The field $\Phi_{mn}$ with $m$ and $n$ being either both even or both odd
positive integers corresponds to the ``degenerate" primary field in the NS
sector with the conformal dimension $\Delta=\Delta(\lambda_{mn})$,
\begin{equation}
\lambda_{mn}=\frac{m b^{-1}+nb}{2}.
\end{equation}
The general form of the descendant operator in the conformal class
$[\Phi_{\Delta}]$ is
\begin{equation}
\mathcal L_{\vec k}\,|\Delta\rangle=
L_{-k_1}\cdots L_{-k_n}G_{-r_1}\cdots G_{-r_m}\Phi_{\Delta},
\label{descendent}
\end{equation}
where $\vec k$ denotes $\{k_i,r_j\}$, which is an ordered set of positive
integers and half-integers correspondingly. The relation $\sum_ik_i+
\sum_jr_j=N$ fixes the particular level in the Verma module corresponding to
the superconformal family $[\Phi_{\Delta}]$. As usual, the Ward identities
restrict the possible coordinate dependence of the correlation functions.
In particular, the two-point functions are completely determined:
\begin{equation}
\langle\mathbf{\Phi}_1(z_1)\mathbf{\Phi}_2(z_2)\rangle\sim
\frac{\delta_{\Delta_1,\Delta_2}}{\vert z_{12}\vert^{4\Delta_1}},
\end{equation}
where $z_{ik}=x_i-x_k+\theta_i\theta_k$. The three-point function depends on
an arbitrary function of the ``odd" superprojective invariant of three
points,
\begin{equation}
\theta_{123}=\frac
{z_{23}\theta_1+z_{31}\theta_2+z_{12}\theta_3-\theta_1\theta_2\theta_3}
{\left(z_{12}z_{13}z_{23}\right)^{1/2}},
\end{equation}
and is presented as
\begin{equation}
\langle\mathbf{\Phi}_1(z_1)\mathbf{\Phi}_2(z_2)
\mathbf{\Phi}_3(z_3)\rangle=\frac{C_{123}+
\vert\theta_{123}\vert^2\widetilde{C}_{123}}
{\vert z_{12}\vert^{2(\Delta_1+\Delta_2-\Delta_3)}\vert
z_{13}\vert^{2(\Delta_1+\Delta_3-\Delta_2)}\vert
z_{23}\vert^{2(\Delta_2+\Delta_3-\Delta_1)}}.
\end{equation}
The ``degenerate" primary field $\Phi_{mn}$ has a singular vector at the
level $N=mn/2$~\cite{Kac}. It is convenient to introduce a ``singular-vector
creation operator" $D_{m,n}$~\cite{SUSYHEM} such that the singular vector
appears when $D_{m,n}$ is applied to $\Phi_{mn}$. We fix the normalization by
taking the coefficient of the leading term to be unity, $D_{mn}=
G_{-1/2}^{mn}+\dots$. The first nontrivial null vector in the NS sector
occurs on the level $N=\frac{3}{2}$:
\begin{equation}\label{NULL-VECTOR1}
D_{13}\Phi_{13}=
\left(L_{-1}G_{-\frac{1}{2}}+b^2G_{-\frac{3}{2}}\right)\Phi_{13}=0.
\end{equation}

\section{Four-point correlation function and conformal blocks}

We consider the four-point correlation function of the primary superfields in
the NS sector of the $N{=}1$ SUSY conformal theory. For the four-point
correlation function, there are three independent superprojective invariants,
one ``even" and two ``odd" (see, e.g.,~\cite{Alvarez}). Using the
superconformal invariance (similarly to the consideration in~\cite{3tMLG}),
we can write
\begin{equation}
\langle{\bf\Phi}_1(z_1){\bf\Phi}_2(z_2){\bf\Phi}_3(z_3)
{\bf\Phi}_4(z_4)\rangle=
|z_{34}|^{2\gamma_{34}}|z_{13}|^{2\gamma_{13}}|z_{23}|^{2\gamma_{23}}
|z_{12}|^{2\gamma_{12}}g(z,\bar z,\tau_1,\bar\tau_1,\tau_2,\bar\tau_2),
\label{anzac}
\end{equation}
where
\begin{equation}
\begin{aligned}
&\gamma_{34}=-2\Delta_4,
\\
&\gamma_{13}=-\Delta_1-\Delta_3+\Delta_4+\Delta_2,
\\
&\gamma_{23}=-\Delta_2-\Delta_3+\Delta_4+\Delta_1,
\\
&\gamma_{12}=-\Delta_4-\Delta_1-\Delta_2+\Delta_3,
\end{aligned}
\end{equation}
and we choose three independent invariants:
\begin{equation}
\begin{aligned}
&z=\frac{z_{41}z_{23}}{z_{43}z_{21}},
\\
&\tau_1=\theta_{213},
\\
&\tau_2=-[z(z-1)]^{1/2}\theta_{214}.
\end{aligned}
\end{equation}
Taking the superprojective invariance into account, we can, for example, fix
\begin{equation}
\begin{aligned}
&\theta_1=0,\qquad\theta_2=0,\qquad\theta_3=R\eta,
\\
&x_1=0,\qquad x_2=1,\qquad x_3=R,\qquad x_4=x,
\end{aligned}
\label{params}
\end{equation}
where $R\to\infty$. The function $g$ is related to the correlation function
of the boson components of supermultiplet~\eqref{superfield}:
\begin{equation}
\langle\Phi_1(0)\Phi_2(1)\Phi_3(\infty)\Phi_4(z)\rangle=
g(z,\bar z,0,0,0,0)=g_0(z,\bar z).
\label{g1234}
\end{equation}
In what follows, we restrict ourself to considering superconformal blocks
contributing to the correlation function $g_0$. Generalizing to the other
components is straightforward.

Similarly to the case of the purely conformal symmetry~\cite{BPZ}, we can
write the $s$-channel expansion for the correlation function $g_0$:
\begin{align}
\langle\Phi_1(x)\Phi_2(0)\Phi_3(1)\Phi_4(\infty)\rangle=
\sum_{\Delta}\bigg[&C_{12}^{\Delta}C_{34}^{\Delta}
F_0(\Delta,\Delta_i,x)F_0(\Delta,\Delta_i,\bar x)
\nonumber\\
&{}+\tilde C_{12}^{\Delta}\tilde C_{34}^{\Delta}
F_1(\Delta,\Delta_i,x) F_1(\Delta,\Delta_i,\bar x)\bigg].
\label{corrfun}
\end{align}
The superconformal blocks $F_0$ and $F_1$ are defined similarly to the
bosonic case (see, e.g.,~\cite{ZZIII} for details):
\begin{align}
&F_0(\Delta,\Delta_i,c,x)=x^{\Delta-\Delta_1-\Delta_2}
\sum^{N\text{ integer}}_{N\ge0}x^N{}_{12}\langle N|N\rangle_{34},
\label{ConfBlockDef1}
\\
&F_1(\Delta,\Delta_i,c,x)=x^{\Delta-\Delta_1-\Delta_2}
\sum^{N\text{ half-integer}}_{N>0}x^N{}_{12}\langle N|N\rangle_{34},
\label{ConfBlockDef2}
\end{align}
where the vector $|N\rangle$ is the $N$th-level descendent contribution of
the intermediate state with the conformal dimension $\Delta$ appearing in the
operator product expansion (OPE) $\Phi(x)\Phi(0)$:
\begin{equation}
[\Phi_1(x)\Phi_2(0)]_{\Delta}=
x^{\Delta-\Delta_1-\Delta_2}\sum_{N=0}^{\infty}x^N |N\rangle_{12},
\end{equation}
where $|N\rangle_{12}=Q_{12}(N,\Delta)|\Delta\rangle$, $Q_{12}(N,\Delta)=
\sum\beta_{12}(\vec k)\mathcal L_{\vec k}^{(N)}$, and $\mathcal L_{\vec k}$
is defined in~\eqref{descendent} (we assume summation over all $N$th-level
descendents). The vectors $|N\rangle_{12}$ depend not only on the conformal
dimension $\Delta$ and central charge $c$ but also on the dimensions of the
fields $\Phi_1$ and $\Phi_2$, and the operator $Q$ is hence supplied with the
appropriate subscript. The vectors $|N\rangle_{12}$ and also the vectors
$|\widetilde{N}\rangle_{12}=\widetilde Q(N,\Delta)|\Delta\rangle$ arising in
the OPE $\Psi_1(x)\Phi_2(0)$,
\begin{equation}
[\Psi_1(x)\Phi_2(0)]_{\Delta}=x^{\Delta-\Delta_1-\Delta_2-\frac{1}{2}}
\sum_{N=0}^{\infty} x^N \widetilde{|N\rangle}_{12},
\end{equation}
are completely determined by the superconformal symmetry. Namely, the
superconformal constraints lead to relations for the vectors of the chain
that grows from the vacuum vector $|\Delta\rangle$,
\begin{equation}
\begin{cases}
G_k|N\rangle_{12}={\widetilde{|N-k \rangle}}_{12},\\
G_k{\widetilde{|N\rangle}}_{12}=
[\Delta+2k\Delta_1-\Delta_2+N-k]|N-k\rangle_{12},\end{cases}
\label{chain}
\end{equation}
for $k>0$. From definition~\eqref{ConfBlockDef1},~\eqref{ConfBlockDef2} and
from the properties of Eqs.~\eqref{chain}, we can deduce that starting from
the level $mn/2$, the functions $F_0$ and $F_1$ have simple poles for
$\Delta=\Delta_{mn}(c)$. Similarly, this relation shows that as functions of
the central charge $c$, the conformal blocks have one simple pole for each
pair of positive integers $m$ and $n$ ($n>1$) at $c=c_{mn}(\Delta)$, where
\begin{align}
&c_{mn}=5+2(T_{mn}+T_{mn}^{-1}),
\label{Cmn}
\\
&T_{mn}=\frac{1-4\Delta-mn+\sqrt{[(mn-1)+4\Delta]^2-(m^2-1)(n^2-1)}}{n^2-1}.
\label{Tmn}
\end{align}
The residues of the functions $F_0$ and $F_1$ at these poles are proportional
to the conformal block functions corresponding to the invariant subclass that
appears on the $mn/2$ level with the highest vector being a new primary field
with the conformal dimension $\Delta_{mn}=\Delta+m n/2$ (see an analogous
consideration in~\cite{AlZ1} and also~\cite{ZZIII}). The singular part is
briefly discussed in the next section.

\section{Singular structure of the chain vectors}

We will consider the singular structure of the chain vectors in more detail
in a subsequent publication. Here, we present the main
results concerning the singularities of the chain vectors introduced
in~\eqref{chain}. For $\Delta\to\Delta_{mn}$,
\begin{align}
&|N=\frac{mn}{2}\rangle\to
\frac{X_{mn}}{\Delta-\Delta_{mn}}D_{mn}|\Delta\rangle,
\\
&|\widetilde{N=\frac{mn}{2}}\rangle\to
\frac{{\bar X}_{mn}}{\Delta-\Delta_{mn}}D_{mn}|\Delta\rangle.
\end{align}
To recover the dependence of the coefficient functions $X_{mn}$ and
$\bar X_{mn}$ on the external dimensions, we observe that for
$\Delta=\Delta_{mn}(c)$, the chain vectors should still be well defined if
certain ``fusion" relations~\cite{BPZ} between $\Delta$ and $\Delta_i$ are
satisfied. Investigating the ``fusion" rules based on analyzing the structure
functions~\cite{Pogosian} leads to the expressions
\begin{align}
&X_{mn}=\begin{cases}\vphantom{\Biggl(}
2^{-\frac{mn}{2}}
\dfrac{P_{mn}(\lambda_1+\lambda_2)P_{mn}(\lambda_1-\lambda_2)}{r'_{mn}},&
m,n\text{ even},\\ \vphantom{\Biggl(}
2^{\frac{1-mn}{2}}
\dfrac{P_{mn}(\lambda_1+\lambda_2)P_{mn}(\lambda_1-\lambda_2)}{r'_{mn}},&
m,n\text{ odd}\end{cases}
\\
&\bar X_{mn}=\begin{cases}\vphantom{\Biggl(}
2^{-\frac{mn}{2}}
\dfrac{\bar P_{mn}(\lambda_1+\lambda_2)\bar P_{mn}(\lambda_1-\lambda_2)}
{r'_{mn}},&m,n\text{ even},\\ \vphantom{\Biggl(}
2^{-\frac{1+mn}{2}}
\dfrac{\bar P_{mn}(\lambda_1+\lambda_2)\bar P_{mn}(\lambda_1-\lambda_2)}
{r'_{mn}},&m,n\text{ odd}.\end{cases}
\end{align}

\pagebreak

\noindent
Here,
\begin{equation}
P_{mn}=\prod_{(r,s)\in[m,n]}(\lambda-\lambda_{rs}),
\end{equation}
where
\begin{equation}
[m,n]=\{1-m:4:m-3,1-n:4:n-3\}\cup\{3-m:4:m-1,3-n:4:n-1\}
\end{equation}
for $m$ and $n$ both even and
\begin{equation}
[m,n]=\{3-m:4:m-3,1-n:4:n-1\}\cup\{1-m:4:m-1,3-n:4:n-3\}
\end{equation}
for $m$ and $n$ both odd;
\begin{equation}
{\bar P}_{mn}=\prod_{(r,s)\in\overline{[m,n]}}(\lambda-\lambda_{rs}),
\end{equation}
where
\begin{equation}
\overline{[m,n]}=\{1-m:4:m-3,3-n:4:n-1\}\cup\{3-m:4:m-1,1-n:4:n-3\}
\end{equation}
for $m$ and $n$ both even and
\begin{equation}
\overline{[m,n]}=\{3-m:4:m-3,3-n:4:n-3\}\cup\{1-m:4:m-1,1-n:4:n-1\}
\end{equation}
for $m$ and $n$ both odd. A simple analysis of~\eqref{chain} shows that the
factor $r_{mn}'$ depends only on $c$ and $\Delta$. It will be clarified
in the subsequent publication that this coefficient is related to the norm of the
corresponding singular vector,
\begin{equation}
\|D_{mn}|\Delta\rangle\|^2=r'_{mn}(\Delta-\Delta_{mn})
\end{equation}
as $\Delta\to\Delta_{mn}$. Taking the results in~\cite{SUSYHEM} into account,
we write
\begin{equation}
r'_{mn}=\frac{r_{mn}}{\lambda_{mn}},
\end{equation}
where
\begin{equation}
r_{mn}=2^{1-mn}\prod_{(k,l)\in[m,n]}(kb^{-1}+lb)
\label{rprod}
\end{equation}
and
\begin{equation}
[m,n]=\{1-m:2:m-1,1-n:2:n-1\}\cup\{2-m:2:m,2-n:2:n\}\setminus(0,0).
\end{equation}
The expressions of the form $a:d:b$ (``from $a$ to $b$ step $d$'') in the
above formulas denote sets of numbers $a,a+d,a+2d,\dots,b$. The symbol
$\{A,B\}$ denotes the set of pairs $(k,l)$ with $k$ and $l$ independently
ranging the sets $A$ and $B$, and $\{A_1,B_1\}\cup\{ A_2,M_2\}$ is the
standard union of two sets. Finally, $\ldots\setminus(0,0)$ means that the
pair $(0,0)$ is excluded.

In the same way, the chain vectors for $N>mn/2$ also have simple poles at
$\Delta=\Delta_{mn}$, and hence
\begin{align}
&|N\ge\frac{mn}{2}\rangle\to\begin{cases}
\dfrac{X_{mn}}{\Delta-\Delta_{mn}}
Q\left(N-\dfrac{mn}{2},\Delta_{mn}+\dfrac{mn}{2}\right)D_{mn}|\Delta\rangle,&
N-\dfrac{mn}{2}\text{ integer},\\
\dfrac{{\bar X}_{mn}}{\Delta-\Delta_{mn}}
Q\left(N-\dfrac{mn}{2},\Delta_{mn}+\dfrac{mn}{2}\right)D_{mn}|\Delta\rangle,&
N-\dfrac{mn}{2}\text{ half-integer},\end{cases}
\\
&|\widetilde{N\ge\frac{mn}{2}}\rangle\to\begin{cases}
\dfrac{{\bar X}_{mn}}{\Delta-\Delta_{mn}}
\widetilde Q\left(N-\dfrac{mn}{2},\Delta_{mn}+\dfrac{mn}{2}\right)
D_{mn}|\Delta\rangle,&N-\dfrac{mn}{2}\text{ integer},\\
\dfrac{X_{mn}}{\Delta-\Delta_{mn}}
\widetilde Q\left(N-\dfrac{mn}{2},\Delta_{mn}+\dfrac{mn}{2}\right)
D_{mn}|\Delta\rangle,&N-\dfrac{mn}{2}\text{ half-integer}.\end{cases}
\end{align}

\section{Recursive relations}

The consideration in the preceding section leads to relations for the
conformal blocks as functions of the internal conformal dimension $\Delta$:
\begin{equation}
\begin{aligned}
&F_0(\Delta\to\Delta_{mn},\Delta_i,c,x)=\begin{cases}
\dfrac{R_{mn}}{\Delta-\Delta_{mn}}F_0(\Delta_{m,-n},\Delta_i,c,x),&
m,n\text{ even},\\
\dfrac{\bar R_{mn}}{\Delta-\Delta_{mn}}F_1(\Delta_{m,-n},\Delta_i,c,x),&
m,n\text{ odd},\end{cases}
\\
&F_1(\Delta\to\Delta_{mn},\Delta_i,c,x)=\begin{cases}
\dfrac{\bar R_{mn}}{\Delta-\Delta_{mn}}F_1(\Delta_{m,-n},\Delta_i,c,x),&
m,n\text{ even},\\
\dfrac{R_{mn}}{\Delta-\Delta_{mn}}F_0(\Delta_{m,-n},\Delta_i,c,x),&
m,n\text{ odd},\end{cases}
\end{aligned}
\end{equation}
where
\begin{equation}
R_{mn}=X_{mn}^{(12)}X_{mn}^{(34)}r_{mn}'\qquad\text{and}\qquad
\bar R_{mn}=\bar X_{mn}^{(12)}\bar X_{mn}^{(34)} r_{mn}'.
\label{Rmn}
\end{equation}
Hence, the residues at the poles of the conformal blocks as functions of the
central charge $c$ are also completely determined:
\begin{equation}
\begin{aligned}
&F_0(\Delta,\Delta_i,c\to c_{mn},x)=\begin{cases}
\dfrac{R_{mn}'}{c-c_{mn}}
F_0\left(\Delta+\dfrac{mn}{2},\Delta_i,c_{mn},x\right),&m,n\text{ even},\\
\dfrac{\bar R_{mn}'}{c-c_{mn}}
F_1\left(\Delta+\dfrac{mn}{2},\Delta_i,c_{mn},x\right),&
m,n\text{ odd}.\end{cases}
\\
&F_1(\Delta,\Delta_i,c\to c_{mn},x)=\begin{cases}
\dfrac{\bar R_{mn}'}{c-c_{mn}}
F_1\left(\Delta+\dfrac{mn}{2},\Delta_i,c_{mn},x\right),&m,n\text{ even},\\
\dfrac{R_{mn}'}{c-c_{mn}}
F_0\left(\Delta+\dfrac{mn}{2},\Delta_i,c_{mn},x\right),&
m,n\text{ odd}.\end{cases}
\end{aligned}
\end{equation}
where $c_{mn}$ is defined in~\eqref{Cmn} and~\eqref{Tmn} and the coefficients
$R_{mn}$ and $R_{mn}'$ are essentially the same and differ only in the change
of variable
\begin{align}
&R_{mn}'=R_{mn}(c_{mn})\cdot
\bigg(\frac{\partial\Delta_{mn}}{\partial c}\bigg)^{-1},
\\
&\frac{\partial\Delta_{mn}}{\partial c}=
\frac{1}{16}\frac{(1-n^2)T_{mn}-(1-m^2)T_{mn}^{-1}}{T_{mn}-T_{mn}^{-1}}.
\end{align}
For $c=\infty$, relations~\eqref{chain} are simplified and can be solved
explicitly. This leads to expressions for the asymptotic values of $F_0$
and $F_1$ in terms of hypergeometric functions:
\begin{equation}
\begin{aligned}
F_0(c\,{\to}\infty)&=f_0(\Delta,\Delta_i,x)=
x^{\Delta-\Delta_1-\Delta_2}
{}_2F_1(\Delta+\Delta_1-\Delta_2,\Delta+\Delta_3-\Delta_4,2\Delta,x),
\\
F_1(c\,{\to}\infty)&=f_1(\Delta,\Delta_i,x)
\\
&=\frac{1}{2\Delta}\,x^{\Delta-\Delta_1-\Delta_2+1/2}
{}_2F_1(\Delta+\Delta_1-\Delta_2+1/2,\Delta+\Delta_3-\Delta_4+1/2,2\Delta+1,x).
\end{aligned}
\end{equation}
It is therefore clear that we can write the following relations for the
conformal blocks $F_0$ and $F_1$:
\begin{align}
F_0(\Delta,\Delta_i,c,x)=f_0(\Delta,\Delta_i,x)&{}+
\sum_{\{m,n\}}^{m,n\text{ even}}\frac{R_{mn}'}{c-c_{mn}}
F_0(\Delta_{m,-n},\Delta_i,c_{mn},x)
\nonumber
\\
&{}+\sum_{\{m,n\}}^{m,n\text{ odd},\,n>1}\frac{\bar R_{mn}'}{c-c_{mn}}
F_1(\Delta_{m,-n},\Delta_i,c_{mn},x),
\label{recur1}
\\
F_1(\Delta,\Delta_i,c,x)=f_1 (\Delta,\Delta_i,x)&{}+
\sum_{\{m,n\}}^{m,n\text{ even}}\frac{\bar R_{mn}'}{c-c_{mn}}
F_1 (\Delta_{m,-n},\Delta_i,c_{mn},x)
\nonumber
\\
&{}+\sum_{\{m,n\}}^{m,n\text{ odd},\,n>1}\frac{R_{mn}'}{c-c_{mn}}
F_0(\Delta_{m,-n},\Delta_i,c_{mn},x).
\label{recur2}
\end{align}
We can expand $F_0$ and $F_1$ in $x$ by iterating Eqs.~\eqref{recur1}
and~\eqref{recur2}:
\begin{equation}
\begin{aligned}
&F_0(\Delta,\Delta_i,c,x)=x^{\Delta-\Delta_1-\Delta_2}
\sum_{k=0}^{\infty}F_0^{(k)}x^k,
\\
&F_1(\Delta,\Delta_i,c,x)=x^{\Delta-\Delta_1-\Delta_2+1/2}
\sum_{k=0}^{\infty}F_1^{(k)}x^k.
\end{aligned}
\end{equation}
For brevity, the dependence on the external dimensions, which is always the
same, is omitted below. Using recursive relations~\eqref{recur1}
and~\eqref{recur2}, we easily find the first few terms of the series
expansions:
\begin{equation}
\begin{aligned}
&F_0^{(0)}=f_0^{(0)}(\Delta),
\\
&F_0^{(1)}=f_0^{(1)}(\Delta),
\\
&F_0^{(2)}=f_0^{(2)}(\Delta)+\frac{\bar R_{13}'}{c-c_{13}}
f_1^{(0)}\left(\Delta+\frac{3}{2}\right)+
\frac{R_{22}'}{c-c_{22}}f_0^{(0)}(\Delta+2),
\\
&F_1^{(0)}=f_1^{(0)}(\Delta),
\\
&F_1^{(1)}=f_1^{(1)}(\Delta)+\frac{\bar R_{13}'}{c-c_{13}}
f_0^{(0)}\left(\Delta+\frac{3}{2}\right),
\\
&F_1^{(2)}=f_1^{(2)}(\Delta)+\frac{R_{13}'}{c-c_{13}}
f_0^{(1)}\left(\Delta+\frac{3}{2}\right)+
\frac{\bar R_{22}'}{c-c_{22}}f_1^{(0)}(\Delta+2)+
\frac{R_{15}'}{c-c_{15}}f_0^{(0)}\left(\Delta+\frac{5}{2}\right).
\end{aligned}
\end{equation}
Using Eqs.~\eqref{Rmn} and also the expressions obtained in the preceding
section, we can write the first coefficients explicitly:
{\allowdisplaybreaks
\begin{align}
F_0^{(0)}={}&1,
\label{F0}
\\
F_0^{(1)}={}&
(\Delta+\Delta_1-\Delta_2)(\Delta+\Delta_3-\Delta_4)
(2\Delta)^{-1},
\\
F_0^{(2)}={}&(\Delta+\Delta_1-\Delta_2)(1+\Delta+\Delta_1-\Delta_2)
(\Delta+\Delta_3-\Delta_4)(1+\Delta+\Delta_3-\Delta_4)
\nonumber
\\*
&{}\times\bigl(4\Delta(1+2\Delta)\bigr)^{-1}
\nonumber
\\*
&{}+(\Delta^2-3(\Delta_1-\Delta_2)^2+2\Delta(\Delta_1+\Delta_2))
\nonumber
\\*
&{}\times(\Delta^2-3(\Delta_3-\Delta_4)^2+2\Delta(\Delta_3+\Delta_4))
\bigl(2\Delta(3+2\Delta)(-3+3 c+16\Delta)\bigr)^{-1}
\nonumber
\\*
&{}+(\Delta_1-2(\Delta_1-\Delta_2)^2+\Delta_2+\Delta(-1+2\Delta_1+2\Delta_2))
\nonumber
\\*
&{}\times
(\Delta_3-2(\Delta_3-\Delta_4)^2+\Delta_4+\Delta(-1+2\Delta_3+2\Delta_4))
\nonumber
\\*
&{}\times\bigl((c+2(-3+c)\Delta+4\Delta^2)(3+4\Delta(2+\Delta))\bigr)^{-1},
\\
F_1^{(0)}={}&(2\Delta)^{-1},
\label{F1}
\\
F_1^{(1)}={}&(1+2\Delta+2\Delta_1-2\Delta_2)
(1+2\Delta+2\Delta_3-2\Delta_4)
\bigl(8\Delta(1+2\Delta)\bigr)^{-1}\nonumber
\\*
&{}+4(\Delta_1-\Delta_2)(\Delta_3-\Delta_4)\bigl((c+2(-3+c)\Delta+4 \Delta^2)\Delta(1+2\Delta)\bigr)^{-1},
\\
F_1^{(2)}={}&128^{-1}
(1+2\Delta+2\Delta_1-2\Delta_2)(3+2\Delta+2\Delta_1-2\Delta_2)
(1+2\Delta+2\Delta_3-2\Delta_4)
\nonumber
\\
&{}\times(3+2\Delta+2\Delta_3-2\Delta_4)
\bigl(\Delta(1+3\Delta+2\Delta^2)\bigr)^{-1}
\nonumber
\\
&{}+(3+2\Delta+2\Delta_1-2\Delta_2)(\Delta_1-\Delta_2)
(3+2\Delta+2\Delta_3-2\Delta_4)(\Delta_3-\Delta_4)
\nonumber
\\
&{}\times\bigl((c+2(-3+c)\Delta+4\Delta^2)(3+4\Delta(2+\Delta))\bigr)^{-1}
\nonumber
\\
&{}+(1+4\Delta_1+4\Delta_2+2(-6(\Delta_1-\Delta_2)^2+
\Delta(-1+2\Delta_1+2\Delta_2)))
\nonumber
\\
&{}\times(\Delta_1,\Delta_2\to\Delta_3,\Delta_4)
\bigl(64\Delta(1+\Delta)(2+\Delta)
(5-11\Delta+2\Delta^2+3c(1+\Delta))\bigr)^{-1}
\nonumber
\\
&{}+(3(-1+4\Delta_1+4\Delta_2)+4(\Delta^2-3(\Delta_1-\Delta_2)^2+
\Delta(1+2\Delta_1+2\Delta_2)))
\nonumber
\\
&{}\times(\Delta_1,\Delta_2\to\Delta_3,\Delta_4)
\bigl(64\Delta(2+\Delta)(3+2\Delta)(-3+3c+16\Delta)\bigr)^{-1}.
\end{align}
}

\section{Differential equations corresponding to the null vector (1,3)}

Because singular vector~\eqref{NULL-VECTOR1} vanishes, the differential
equation
\begin{equation}
\left\lbrace-b^{-2}\partial_4D_4+
\sum_{i=1}^3\left\lbrace\frac{2\Delta_i}{z_{4i}^2}\theta_{4i}+
\frac{1}{z_{4i}}\bigg[2\theta_{4i}\partial_i-D_i\bigg]\right\rbrace
\right\rbrace\langle{\bf\Phi}_1(z_1){\bf\Phi}_2(z_2){\bf\Phi}_3(z_3)
{\bf\Phi}_{13}(z_4)\rangle=0
\label{difur0}
\end{equation}
holds, where
\begin{equation}
D=\partial_{\theta}-\theta\partial_x.
\end{equation}
With~\eqref{anzac}, this equation reduces to the differential equation for
$g(z,\bar z,\tau_1,\bar\tau_1,\tau_2,\bar\tau_2)$
\begin{align}
\bigg[-b^{-2}\partial_4D_4-b^{-2}\frac{\gamma_{34}}{z_{34}}
\bigg(\theta_{43}\partial_4-D_4\bigg)+
\sum_{i=1}^3\bigg[\frac{2\Delta_i}{z_{4i}^2}\theta_{4i}+
\frac{1}{z_{4i}}\bigg(2\theta_{4i}\partial_i-D_i\bigg)\bigg]&
\nonumber
\\
{}+\frac{\gamma_{12}(\theta_{41}+\theta_{42})}{z_{41}z_{42}}+
\frac{\gamma_{13}(\theta_{41}+\theta_{43})}{z_{41}z_{43}}+
\frac{\gamma_{23}(\theta_{42}+\theta_{43})}{z_{42}z_{43}}+
\frac{\gamma_{34}\theta_{43}}{z_{43}^2}&\bigg]g=0,
\label{difur}
\end{align}
In accordance with~\eqref{g1234}, we present $g=g_0(z)+g_1(z)\tau_1+
g_2(z)\tau_2+g_3(z)\tau_1\tau_2$ (also keeping in mind the antiholomorphic
dependence). Equation~\eqref{difur} splits into two independent systems of
ordinary differential equations for $g_i$. For the purpose of this paper, we
explicitly write the system for $g_0$ and $g_3$:
\begin{align}
&-b^{-2}zg_0''+\frac{3z-2}{z-1}g_0'+b^{-2}g_3'+
\bigg[\frac{\gamma_{13}}{z}+\frac{\gamma_{23}}{z-1}\bigg]g_0+
\frac{1-2z}{z(z-1)}g_3=0,
\label{ur1}
\\
&b^{-2}g_0''+\frac{1-3z}{z(z-1)}g_0'+\bigg[\frac{2\Delta_1}{z^2}+
\frac{2\Delta_2}{(z-1)^2}+\frac{2\gamma_{12}}{z(z-1)}\bigg]g_0+
\frac{1}{z(z-1)}g_3=0.
\label{ur2}
\end{align}
Three independent solutions for $g_0$ with a diagonal monodromy near $x=0$
are just the $s$-channel conformal blocks~(\ref{ConfBlockDef1},\ref{ConfBlockDef2}) with the special parameter choices
$\Delta_1=\Delta_{13}$, $\Delta_2=\Delta(\lambda_1)$, $\Delta_3=
\Delta(\lambda_2)$, $\Delta_4=\Delta(\lambda_3)$, and $\Delta=
\Delta^{(\pm)}=\Delta(\lambda_1\pm b)$ or $\Delta=\Delta^{(0)}=
\Delta(\lambda_1)$,
\begin{align}
&g_0^{(\pm)}=x^{\Delta-\Delta_1-\Delta_2}\sum_{n=0}^{\infty}A_n^{(\pm)}x^n=
F_0(\Delta^{(\pm)},\Delta_i,c,x),
\\
&g_0^{(0)}=x^{\Delta-\Delta_1-\Delta_2+1/2}\sum_{n=0}^{\infty}A_n^{(0)}x^n=
F_1(\Delta^{(0)},\Delta_i,c,x),
\end{align}
corresponding to the overall normalization $A_0^{(\pm)}=1,A_0^{(0)}=
1/(2\Delta(\lambda_1))$. The first terms in the series expansion can be
easily found by substituting these expansions in differential
equations~\eqref{ur1} and~\eqref{ur2} and solving the recursive relations for
the coefficients order by order (we did this using Mathematica),
\begin{align}
A_0^{(+)}={}&1,
\\
A_1^{(+)}={}&(1+2b^2-3b^4-8b^3\lambda_1-4b^2\lambda_1^2-4b^2\lambda_2^2+
4b^2\lambda_3^2)\bigl(4(-1+b^2+2b\lambda_1)\bigr)^{-1},
\\
A_2^{(+)}={}&(21+4b^2-90b^4+84b^6-19b^8-62b\lambda_1-152b^3\lambda_1+
380b^5\lambda_1-184b^7\lambda_1+18b^9\lambda_1
\nonumber
\\
&{}+8b^2\lambda_1^2+432b^4\lambda_1^2-472b^6\lambda_1^2+96b^8\lambda_1^2+
112b^3\lambda_1^3-416b^5\lambda_1^3+176b^7\lambda_1^3-112b^4\lambda_1^4\qquad
\nonumber
\\
&{}+128b^6\lambda_1^4+32b^5\lambda_1^5-72b^2\lambda_2^2+176b^4\lambda_2^2-
40b^6\lambda_2^2+240b^3\lambda_1\lambda_2^2-288b^5\lambda_1\lambda_2^2
\nonumber
\\
&{}+48b^7\lambda_1\lambda_2^2-224b^4\lambda_1^2\lambda_2^2+
128b^6\lambda_1^2\lambda_2^2+64b^5\lambda_1^3\lambda_2^2-
48b^4\lambda_2^4+32b^5\lambda_1\lambda_2^4
\nonumber
\\
&{}+24b^2\lambda_3^2-144b^4\lambda_3^2+56b^6\lambda_3^2-
112b^3\lambda_1\lambda_3^2+288b^5\lambda_1\lambda_3^2-
48b^7\lambda_1\lambda_3^2+160b^4\lambda_1^2\lambda_3^2
\nonumber
\\
&{}-128b^6\lambda_1^2\lambda_3^2-64b^5\lambda_1^3\lambda_3^2+
96b^4\lambda _2^2\lambda_3^2-64b^5\lambda_1\lambda_2^2\lambda_3^2-
48b^4\lambda_3^4+32b^5\lambda_1\lambda_3^4)
\nonumber
\\
&{}\times\bigl(64(-1+b\lambda_1)(-3+b^2+2b\lambda_1)
(-1+b^2+2b\lambda_1)\bigr)^{-1},
\\
A_0^{(0)}={}&4b^2\bigl((1+b^2-2b\lambda_1)(1+b^2+2b\lambda_1)\bigr)^{-1},
\\
A_1^{(0)}={}&(-2b^4(9+6b^2+b^4-4b^2\lambda_1^2-8\lambda_2^2-4b^2\lambda_2^2+
8\lambda_3^2+4b^2\lambda_3^2))
\nonumber
\\
&{}\times\bigl((1+b^2-2b\lambda_1)(3+b^2-2b\lambda_1)(1+b^2+2b\lambda_1)
(3+b^2+2b\lambda_1)\bigr)^{-1},
\\
A_2^{(0)}={}&(b^2(-72-534b^2-287b^4+48b^6+62b^8+14b^{10}+b^{12}+
32 b^2\lambda_1^2+288b^4\lambda_1^2
\nonumber
\\
&{}-8b^6\lambda_1^2-48b^8\lambda_1^2-8b^{10}\lambda_1^2-32b^6\lambda_1^4+
16b^8\lambda_1^4+544b^2\lambda_2^2+128b^4\lambda_2^2-200b^6\lambda_2^2
\nonumber
\\
&{}-80b^8\lambda_2^2-8b^{10}\lambda_2^2-128b^4\lambda_1^2\lambda_2^2+
32b^8\lambda_1^2\lambda_2^2+128b^4\lambda_2^4+96b^6\lambda_2^4+
16b^8\lambda_2^4
\nonumber
\\
&{}-256b^2\lambda_3^2+208b^4\lambda_3^2+328b^6\lambda_3^2+96b^8\lambda_3^2+
8b^{10}\lambda_3^2-64b^6\lambda_1^2\lambda_3^2-32b^8\lambda_1^2\lambda_3^2
\nonumber
\\
&{}-256b^4\lambda_2^2\lambda_3^2-192b^6\lambda_2^2\lambda_3^2-
32b^8\lambda_2^2\lambda_3^2+128b^4\lambda_3^4+96b^6\lambda_3^4+
16b^8\lambda_3^4))
\nonumber
\\
&{}\times\bigl(2(1+b^2-2b\lambda_1)
(3+b^2-2b\lambda_1)(5+b^2-2b\lambda_1)
\nonumber
\\
&{}\times(1+b^2+2b\lambda_1)(3+b^2+2b\lambda_1)
(5+b^2+2b\lambda_1)\bigr)^{-1}.
\end{align}
These expressions coincide with the expressions for the corresponding
coefficients~\eqref{F0} and~\eqref{F1} calculated for the same parameter
choices and thus confirm recursive relations~\eqref{recur1}
and~\eqref{recur2}.

\section{Conclusion}

Recursive relations~\eqref{recur1} and~\eqref{recur2} for the superconformal
block functions together with the supporting verification in Sec.~6 is our
main result in this paper. These relations allow efficiently evaluating the
superconformal blocks for any parameter values. They are appropriate for
numerically calculating the four-point correlation functions of the primary
fields in the NS sector in general and of the degenerate primary fields in
particular. We note that for the four-point conformal blocks (and also
superconformal blocks), there exists~\cite{AlZ1,AlZ2} the so-called
$q$-representation of the recursive relations, which is much more favorable
because it converges in the whole complex plane with three punctures.

Representations like~\eqref{corrfun} can also be written for higher
multipoint correlation functions. They involve the multipoint conformal
blocks, which are much more complicated than the four-point blocks.
We hope that the consideration in this paper will help to approach the
extremely interesting question of generalizing the recursive relations to
higher multipoint conformal blocks and in particular the question of the
corresponding generalized $q$-representations.

\section{Acknowledgments}
The author is grateful to A.~Belavin and Al.~Zamolodchikov for the useful
discussions and for the possibility to use our results before publication. Special gratitude is extended to A.~Delfino and G.~Mussardo for encouraging the interest in this work.
The author also thanks SISSA for the warm hospitality during the period in which this work was done and INFN for the partial financial support.

\section{Note Added}
After this paper was sent to arXiv I have learned that the recursive relations for the superconformal block functions apparantly equivalent to~\eqref{recur1} and~\eqref{recur2} were proposed simultaneously in~\cite{Leszek}. I am grateful to L.~Hadasz for bringing the paper~\cite{Leszek} to my attention.

\end{document}